\documentclass[journal]{IEEEtran}
\usepackage{amsmath,amsfonts,amssymb}
\usepackage{algorithmic}
\usepackage{algorithm}
\usepackage{array}
\usepackage[caption=false,font=normalsize,labelfont=sf,textfont=sf]{subfig}
\usepackage{textcomp}
\usepackage{stfloats}
\usepackage{url}
\usepackage{verbatim}
\usepackage{graphicx}
\usepackage{cite}
\usepackage{xcolor}
\usepackage{setspace}
\usepackage{booktabs}
\usepackage{multirow}
 \usepackage{arydshln}

\newcommand{\bx}{\mathbf{x}}
\newcommand{\bp}{\mathbf{p}}
\newcommand{\by}{\mathbf{y}}

\newcommand{\bz}{\mathbf{z}}
\newcommand{\bZ}{\mathbf{Z}}
\newcommand{\cZ}{\mathcal{Z}}

\newcommand{\bU}{\mathbf{U}}

\newcommand{\bW}{\mathbf{W}}
\newcommand{\R}{\mathbb{R}}

\renewcommand{\L}{\mathcal{L}}

\begin{document}

\title{Fully Differentiable Neural Forced Alignment via Soft Dynamic Programming}

\author{Rotem Rousso, Eyal Cohen and Joseph Keshet,~\IEEEmembership{Senior Member,~IEEE}
\thanks{Manuscript received February 3, 2026; revised \today.}}

\markboth{Journal of \LaTeX\ Class Files,~Vol.~15, No.~2, February~2026}%
{Rousso \MakeLowercase{\textit{et al.}}: Fully Differentiable Neural Forced Aligner}

\maketitle


\begin{abstract}
Recent advances in sequence modeling have significantly improved ASR systems, bringing them close to human-level recognition accuracy and enhancing robustness across diverse acoustic conditions and languages. In contrast, Forced Alignment has not experienced comparable progress, and traditional HMM-GMM frameworks remain widely adopted and highly competitive. 

To address this gap, we propose an end-to-end, fully differentiable neural architecture specifically designed for phoneme alignment. The model consists of an encoder that processes the input signal and a decoder that produces alignment decisions. The encoder is structured into two complementary branches: one dedicated to phoneme identity verification and the other to phoneme boundary detection. The decoder is implemented as a trainable module based on differentiable soft dynamic programming. The entire system is optimized end-to-end using a novel contrastive loss that encourages clear separation between steady-state phoneme regions and transition boundaries.

The proposed approach outperforms the current state of the art in phoneme alignment on hand-annotated English benchmarks, achieves strong word-level generalization results, and demonstrates generalization on unseen languages.
\end{abstract}


\section{Introduction}
\IEEEPARstart{F}{orced} alignment (FA) aims to determine the temporal boundaries of phonemes in a speech signal given a known transcript. Accurate phoneme alignment is essential for numerous speech processing tasks. It supports a wide range of applications, including speech synthesis, corpus annotation, pronunciation modeling, prosody analysis, and language learning technologies. 


Despite substantial progress in neural automatic speech recognition (ASR), classical statistical aligners remain highly competitive and often outperform modern neural systems in boundary detection accuracy.

Traditional FA systems rely on Gaussian Mixure Model-Hidden Markov Model (GMM-HMM) acoustic models combined with Viterbi decoding. Toolkits such as the Montreal Forced Aligner (MFA) \cite{McAuliffe17-MFA} are widely used due to their ease of use and precise temporal modeling. These systems model phoneme transitions via HMM state transition probabilities and compute frame-level acoustic likelihoods with GMMs, enabling reliable boundary estimation. However, they typically depend on pronunciation dictionaries and grapheme-to-phoneme (G2P) conversion \cite{tang2012discriminative}. This reliance introduces an inherent limitation, as canonical pronunciations generated by G2P models may differ from the actual acoustic realizations in spontaneous speech \cite{greenberg1996insights}. As a result, alignment accuracy may degrade when lexical pronunciations do not match the spoken signal.


Recent advances in end-to-end speech processing have shifted the research focus toward neural architectures capable of learning rich acoustic representations directly from raw waveforms. Large-scale end-to-end models such as wav2vec2.0\cite{wav2vec2}, HuBERT\cite{Hsu21-HUB}, and Whisper \cite{Radford22-RSR} learn powerful acoustic representations and achieve state-of-the-art transcription accuracy. Nevertheless, these architectures are primarily optimized for recognition rather than for precise temporal localization. Consequently, alignment timestamps are often obtained as a byproduct of decoding and may lack the temporal precision required for phoneme-level or word-level segmentation. This objective mismatch explains why classical HMM-based aligners remain competitive for forced alignment tasks despite the progress of neural ASR systems \cite{Rousso24-FACOMP}.

Several recent approaches attempt to adapt neural ASR models for alignment. For example, WhisperX \cite{Bain23-WhisperX} combines a Whisper-based representation, trained on approximately 680k hours to optimize the cross-entropy objective over tokens, with a wav2vec2.0 model trained as a phoneme classifier to estimate word-level timestamps. Similarly, the Massively Multilingual Speech Model (MMS) \cite{pratap2024scaling} follows the wav2vec2.0 architecture but is trained on over 1,100 languages and approximately 500k hours of speech. Another model is the NVIDIA-Canary-1B \cite{nvidia-canary}, which provides segment-level timestamps for ASR using auxilary CTC model. While effective for transcription-oriented applications, these systems do not explicitly optimize boundary precision and perform at the word level. Many neural alignment approaches rely on the Connectionist Temporal Classification (CTC) \cite{Graves06-CTC} objective, which marginalizes over possible alignments and therefore does not directly encourage accurate boundary localization.
In parallel, recent work on unsupervised speech segmentation has shown that acoustic contrast between adjacent frames provides strong cues for phoneme boundary detection. Contrastive approaches such as UnSupSeg \cite{Kreuk20-SCL} learn representations that highlight phoneme transitions without relying on labeled data. However, such methods operate without access to a known transcript and therefore cannot directly address the forced alignment challenge, nor are they built to leverage the knowledge we have on the transcript.


These observations highlight the need for alignment models that explicitly emphasize phoneme transitions while allowing direct optimization of the alignment process. To address this challenge, this work introduces a neural forced-alignment system specifically designed for phoneme boundary estimation. The proposed approach combines contrastive representation learning with a contextual modeling branch and a differentiable dynamic programming decoder, enabling end-to-end optimization of the alignment process. The representation encoder learns boundary-sensitive acoustic features by separating intra-phoneme regions from phoneme transition regions in the latent space, while the contextual encoder aggregates information across neighboring frames to produce temporally consistent phoneme-level posterior probabilities. Together, the latent features from both encoders provide rich acoustic and linguistic information that the soft dynamic programming (Soft-DP) decoder leverages to predict precise phoneme boundaries while preserving gradient flow through the alignment procedure. Our implementation and trained models, with a demo application web page for audio samples, are available here.\footnote{https://github.com/MLSpeech/FALCON/}

Experimental results demonstrate that the proposed method achieves improved phoneme boundary accuracy compared to both classical HMM-GMM-based aligners and recent neural alignment approaches on hand-annotated benchmarks. In addition, the model exhibits strong generalization to word-level alignment and unseen languages.
The main contributions of this work are summarized as follows:
\begin{enumerate}
\item A neural forced alignment architecture that performs on the phoneme-level, and is capable of generalizing to word-level alignment and unseen languages without further training.
\item A contrastive loss (MNCE) that is designed to emphasize acoustic differences between intra-phoneme regions and phoneme transition regions.
\item A fully-differentiable alignment scheme based on soft dynamic programming that enables end-to-end optimization of the alignment process.
\end{enumerate}

\section{Method}

Let $\bx \in \mathcal{X} \subset \R^{T}$ denote a speech waveform composed of $T$ samples. Denote by $\bp = (p_1, \dots, p_N)$ the phoneme sequence pronounced in the waveform, with length $N$, where $p_i \in \mathcal{P}$ for $1 \le i \le N$. Here, $\mathcal{P}$ denotes the set of phonemes in the language, and $|\mathcal{P}|$ its cardinality. We further assume the existence of an alignment vector indicating the start time of each phoneme, $\by^* = (y_1, \ldots, y_N)$, where $y_i \in [1, T]$ for $1 \le i \le N$. The objective of the phoneme alignment task is, given a speech signal and its corresponding phoneme sequence, to accurately estimate the alignment sequence $\hat{\by} = (\hat{y}_1, \dots, \hat{y}_N)$.

Our system comprises three building blocks. The first component is a \emph{representation encoder} $f_\theta$ parameterized by $\theta$, which maps the input speech waveform to a latent representation, $\bZ = f_\theta(\bx)$. The encoder receives the speech waveform $\bx$ as input and generates the representation $\bZ\in\mathcal{Z} \subset \R^{D \times T_s}$, consisting of a sequence of $T_s$ vectors, each of dimension $D$. We will use the term \emph{frames} to denote the time span of each of the $T_s$ vectors. The encoder's temporal resolution determines the alignment resolution of the downstream process and, consequently, the final predicted alignment. 

The second building block is a \emph{context encoder} $g_\psi$ with parameters $\psi$. It takes as input the speech representation $\bZ$ and the phoneme sequence $\bp$, and computes $\bU = g_\psi(\bZ, \bp)$. The resulting matrix $\bU \in \R^{|\mathcal{P}| \times T_s}$ provides, for each time frame, a probability distribution over the $|\mathcal{P}|$ phonemes. 

The final component is the \emph{decoder}, $h_{\bW}$, parameterized by $\bW$. It takes as input the latent representation $\bZ$, the frame-level phoneme distributions $\bU$, and the phoneme sequence $\bp$, and produces the predicted alignment vector $\hat{\by}^* = h_{\bW}(\bZ, \bU, \bp)$. The decoder computes this alignment using a learned soft dynamic programming procedure, as described below.

\begin{figure}[h]
    \centering
    \includegraphics[width=\columnwidth]{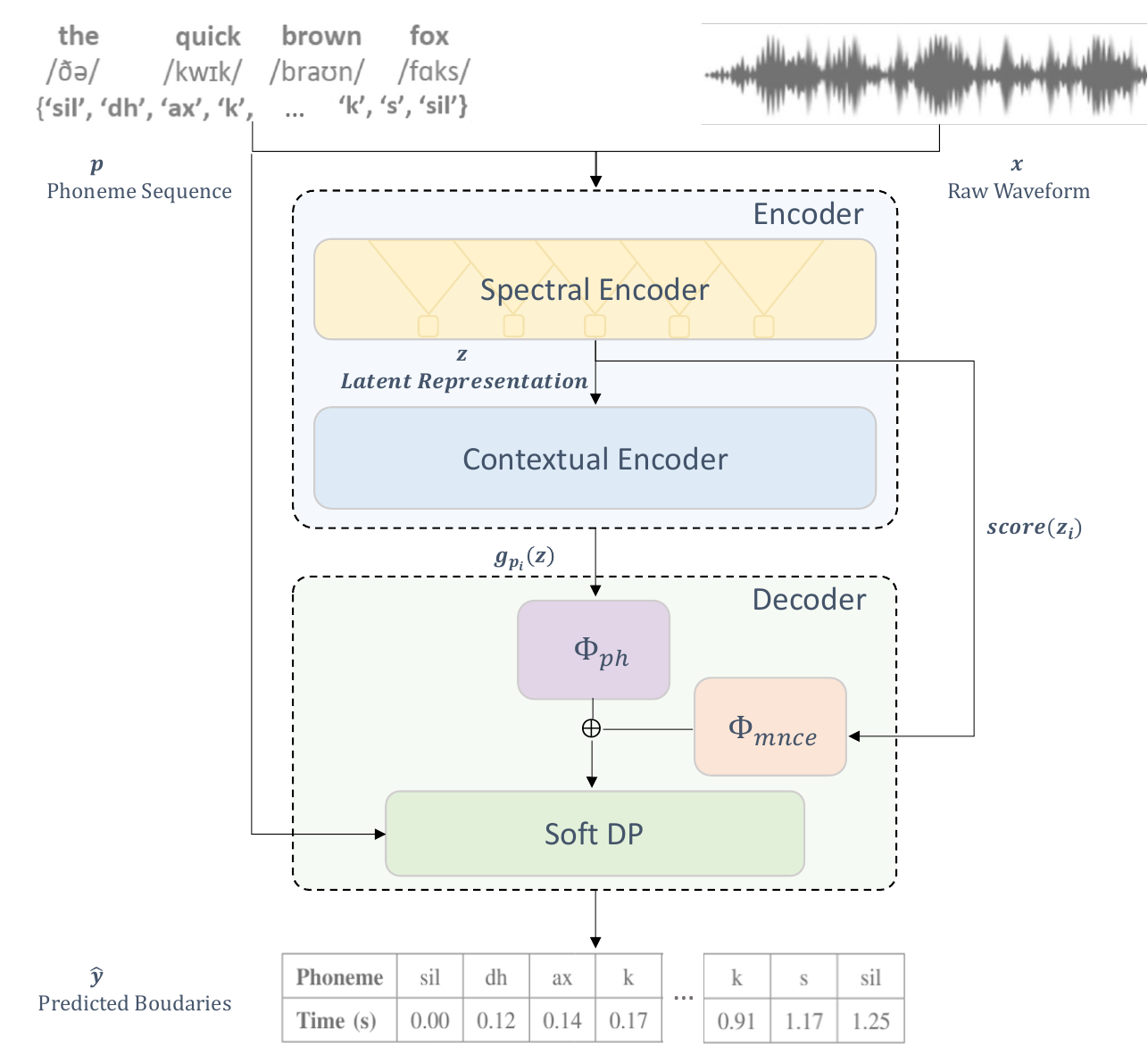}
    \caption{Overview of the proposed phoneme alignment system. The representation encoder extracts latent features from the input waveform, the context encoder produces frame-wise phoneme probabilities, and the decoder predicts the phoneme alignment using differentiable dynamic programming.}
    \label{fig:gantt_dcaf}
\end{figure}

The workflow begins with the representation encoder, which converts the input waveform $\bx$ into a sequence of latent features $\bZ$. The features, together with the phoneme sequence, are processed by the context encoder to generate frame-wise phoneme probabilities. Finally, the decoder predicts the alignment using a differentiable dynamic programming procedure. Each component is trained jointly in an end-to-end manner using the combined objective. The overall architecture of the system is illustrated in Fig. \ref{fig:gantt_dcaf}. 

Overall, given a training set of $M$ examples $S = \{\bx_m, \bp_m, \by_m \}_{m=1}^M$, we aim to minimize a combined objective function, which is composed of three loss functions, that will be explained in detail in the next subsections. The first loss function $\L_{\text{MNCE}}$ encourages the encoder to distinguish between intra-phoneme frames and boundary frames within the latent representation $\bz$. The second loss, $\L_{\text{CE}}$, supervises the context encoder to produce accurate frame-wise phoneme probabilities. The third loss, $\L_{\text{SoftDP}}$, guides the decoder to predict precise phoneme boundaries via differentiable dynamic programming. Our overall loss is a sum of all the three loss functions, where $\L_{\text{CE}}$  is weighted by $\eta$ and the Soft-DP loss,  $\L_{\text{-SoftDP}})$ is weighted by $\mu$, where these weights are found on an held-out set as described in the experimental results.

\section{The Encoder}

The architectural paradigm comprising a representation encoder $f_\theta$ and a context encoder $g_\psi$ has been demonstrated to be effective in prior work \cite{oord2018representation,schneider2019wav2vec}. In this study, we adapt this framework to the task of alignment, with appropriate modifications to its original formulation.

\subsection{Representation encoder}

Building on the recent success of contrastive learning approaches \cite{wav2vec2,Kreuk20-SCL}, we introduce a representation encoder that learns an encoding function $f_\theta: \mathcal{X} \rightarrow \mathcal{Z}$. The proposed objective is formulated to improve the encoder’s ability to discriminate between frames corresponding to inter-phoneme transitions and those arising within intra-phoneme regions. We refer to this objective as Modified Noise Contrastive Estimation (\text{MNCE}), which adapts and refines prior contrastive frameworks grounded in temporal locality \cite{oord2018representation, Kreuk20-SCL} to explicitly support accurate alignment.

Recall that $\bZ=(\bz_1,\ldots,\bz_\tau,\ldots,\bz_{T_s})$ denotes a sequence of frame-level representations, $\bp=(p_1, p_2, \dots, p_K)$ the corresponding phoneme sequence and $\by^*=(y_1, y_2, \dots, y_K)$ their start-times. We define a mapping function $\pi: \{1, \dots, T\} \rightarrow \{1, \dots, K\}$ that assigns each frame index $\tau$ to a phoneme index $i = \pi(\tau)$, where $\pi(\tau) = i$ whenever $y_{i} \leq \tau < y_{i+1}$. This mapping induces a partition of the temporal axis into contiguous segments, where all frames $\tau$ satisfying $\pi(\tau)=i$ are associated with phoneme $p_i$. 

Using the mapping $\pi$ and the duration of each phoneme,  $l_i = y_{i+1} - y_{i}$, we define two subsets of frame indices for each phoneme $p_i$. The positive set $\mathcal{Z}^+_i$ consists of frames lying within the central region of the phoneme segment, excluding frames near the boundaries:
$$\mathcal{Z}^+_i = \bigl\{\bz_\tau \;\big|\; \pi(\tau) = i \;\text{ and }\; y_{i} + 0.25\,l_i \;\leq\; \tau \;\leq\; y_{i} + 0.75\,l_i\bigr\}.$$
The negative set $\mathcal{Z}_i^-$ consists of frames in the immediate vicinity of any phoneme boundary:
$$\mathcal{Z}^-_i = \bigl\{\bz_\tau \;\big|\; \pi(\tau) = i \;\text{ and }\;  |\tau - y_i| \leq \delta\bigr\},$$
where $\delta > 0$ is a half-window parameter controlling the width of the boundary region. By construction, $\mathcal{Z}^+_i$ captures stable intra-phoneme frames, while $\mathcal{Z}^-_i$ captures transitional frames at phoneme boundaries. Note that the two sets are disjoint by design, provided $\delta < 0.2\,l_i$ for all $i$.




The MNCE loss for a single frame $\bz_\tau$ is formulated to contrast distinct temporal regions. Specifically, the objective seeks to maximize the similarity between the $\bz_\tau$ frame and its positive samples set, which represent the stable intra-phoneme region, while minimize the similarity between the $\bz_\tau$ and its negative samples set drawn from the transition boundary region. 
\begin{equation}
\tilde{\mathcal{L}}_{\text{MNCE}}(\bz_\tau, \cZ_i^{-}, \cZ_i^{+}) 
=
- \log 
\frac{
(\sum_{\bz^+_{k} \in \cZ_i^{+}} 
\exp \left(s(\bz_\tau, \bz^+_{k}) \right))^\alpha 
}{
(\sum_{\bz^-_{j} \in\cZ_i^{-}} 
\exp \left(s(\bz_\tau, \bz^-_{j}) \right))^{1-\alpha}
}.
\end{equation}
where $\alpha$ is a learnable parameter that weights the relative importance of positive and negative samples, and $s(\cdot,\cdot)$ denotes the cosine similarity score. The total objective for a training set $S = \{\bx_m, \bp_m, \by_m \}_{m=1}^M$, is obtained by summing the loss over all frames in all sequences
\begin{equation}
\mathcal{L}_{\text{MNCE}} = \sum_{{\bx,\bp,\by}\in S} \sum_{i=1}^{N}\sum_{\tau=y_i}^{y_{i+1}-1} \tilde{\mathcal{L}}_{\text{MNCE}}(\bz_\tau, \cZ_i^{-}, \cZ_i^{+}) 
\end{equation}

\begin{figure}[h]
    \centering
    \includegraphics[width=\columnwidth]{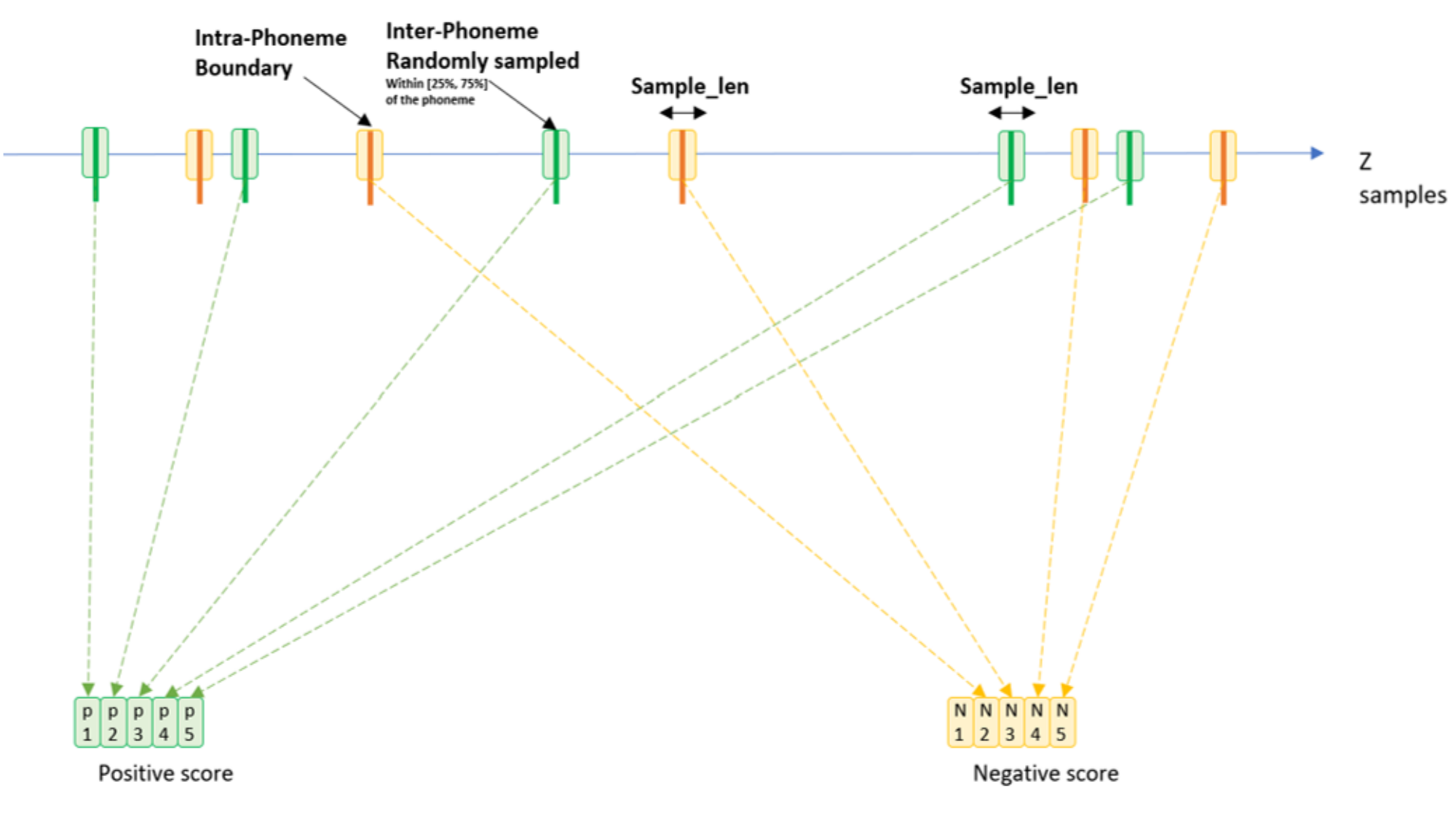}
    \caption{Sampling procedure for the representation encoder. Positive samples (green) are drawn from the steady-state region of the anchor's corresponding phoneme $p_{\pi(\tau)}$, while negative samples (yellow) are sampled from the transition region surrounding the boundary $y_{\pi(\tau)}$.}
    \label{fig: contrastive sampling}
\end{figure}

Fig. \ref{fig: contrastive sampling} illustrates the sampling strategy for the representation encoder. For every frame $\bz_\tau$, the positive samples are restricted to the steady-state region of the same phoneme $p_{\pi(\tau)}$, while the negative samples are concentrated around the boundary $y_{\pi(\tau)}$.

\begin{figure}[h]
    \centering
    \includegraphics[height=3in, width=3.5in]{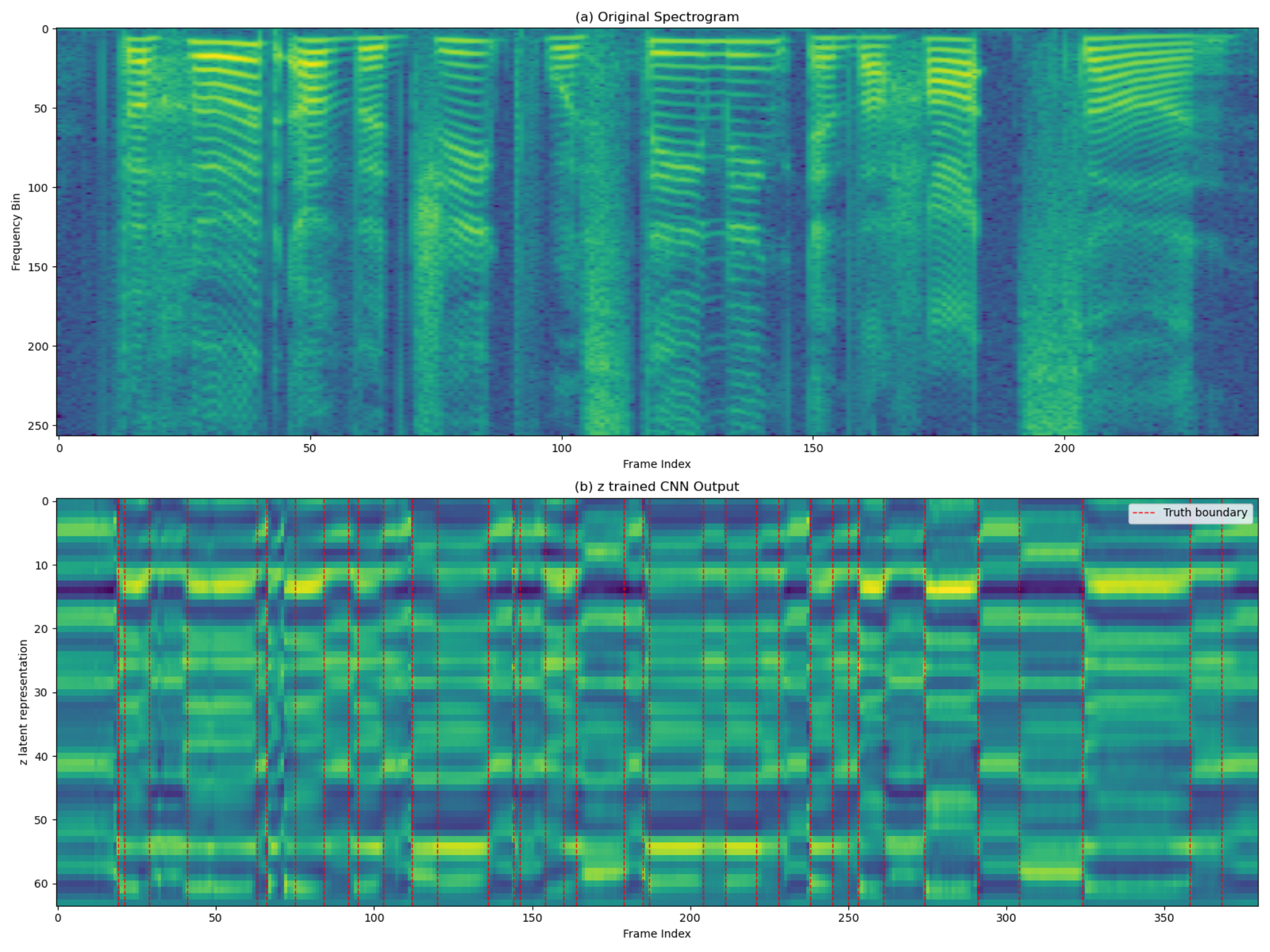}
    \caption{Visualization of the learned latent space: (a) the original spectrogram; (b) the resulting representation $\bz$. Red dashed lines indicate ground truth phoneme boundaries $\by$}
    \label{fig:z_cnn_out}
\end{figure}

As shown in Fig. \ref{fig:z_cnn_out}, the representation $\bz$ learned by the encoder effectively highlights phoneme boundaries. Red dashed lines indicating the ground-truth segmentation. This visualization demonstrates how the objective encourages the model to produce frame-level features that distinguish between the phonemes' transition points.

\subsection{Context encoder}

The context encoder function $g_\psi: \mathcal{Z} \times \mathcal{P} \rightarrow \bU$ is optimized to validate for every frame in $\bZ$ its posterior distribution over all phonemes in the language $\mathcal{P}$. 
While the representation encoder $f_\theta$ captures local acoustic patterns, it is not explicitly optimized for longer-range phonetic temporal dependencies. 
Similar to contextual modules used in CPC \cite{oord2018representation} and wav2vec \cite{schneider2019wav2vec}, the role of $g_\psi$ is to aggregate information across neighboring frames, producing context-aware phoneme probability distributions that capture dependencies that cannot be modeled by local convolutional representation alone. The output of $g_\psi$ is a sequence of phoneme phoneme-level posterior probabilities $\bU \in \mathbb{R}^{|\mathcal{P}| \times T_s}$, where each column represents a categorical distribution over the phoneme set $\mathcal{P}$ for the corresponding frame.


The context encoder is trained using a frame-wise cross-entropy objective. For each frame $\bz_\tau$, the model predicts the probability of the ground truth phoneme $p_i \in \bp$, where $i=\pi(\tau)$ is the phoneme index from the input phoneme sequence assigned to that frame.
The loss is defined as
\begin{equation}
    \mathcal{L}_{\text{CE}}(\bZ,\bp) =  -\sum_{\tau=1}^{T_s} \log P_\psi(p_{\pi(\tau)}\mid \bz_\tau)
\end{equation}
where $P_\psi(p_{\pi(\tau)}\mid \bz_\tau)$ is the posterior probability of the correct phoneme class at time $\tau$.





This objective encourages the model to learn temporally consistent phoneme predictions, which serves as a strong linguistic feature for the decoder to evaluate candidate boundary frames.
Fig. \ref{fig:bilstm_probs} shows the frame-wise phoneme probability map $\bU$ produced by the contextual encoder.


\section{The Decoder}
The final component of our scheme is the alignment decoder, a learnable, differentiable soft-dynamic programming (soft-DP) module that refines alignment decisions. The objective of the decoder is to predict a sequence of $N$ alignments (start times) $\hat{\by}^*$ corresponding to $N$ input phonemes $\bp$. The decoder is built as a decision module based on features $\phi_1 , \phi_2 $ derived from the raw boundary probabilities and the latent spectral representation produced by the encoder.

The decoder takes as input the predicted frame-wise probability map $\bU = g_\psi(\bZ, \bp)$, the latent representation $\bZ$, the input phoneme sequence $\bp$, and the phoneme boundary candidate sequence $\tilde{\textit{\textbf{y}}}$. Formally, the decoder is defined as
\begin{equation}
\hat{\mathbf{y}}^{*} = \arg\max_{\tilde{\textit{\textbf{y}}}}  h_{\bW} (\mathbf{U}, \bZ, \mathbf{p},\tilde{\textit{\textbf{y}}}).
\end{equation}

Following Keshet et al. \cite{Keshet07-LMA}, we model the decoder as a linear combination of two feature functions $\{\phi_n\}$. In dynamic programming recursion, for each step, we denote $\tilde{\textit{\textbf{y}}}$ as the set of current boundary alignment candidates examined. Each feature function evaluates the proposed candidate alignment $\tilde{\textit{\textbf{y}}}$, with higher scores indicating a well-placed alignment and lower scores indicating a poorly aligned one. Formally, for the frames $\tilde{\textit{\textbf{y}}}_{i} , \tilde{\textit{\textbf{y}}}_{i+1}$ as boundary candidates for the input phoneme $p_i$, the local alignment score is formulated as follows 
\begin{equation}
D_{i,\tilde{\textit{\textbf{y}}}_{i+1},\tilde{\textit{\textbf{y}}}_{i}} = W_1\cdot \phi_1(\bz,\tilde{\textit{\textbf{y}}}_{i}) + W_2\cdot \phi_2(\bU, p_i,\tilde{\textit{\textbf{y}}}_{i},\tilde{\textit{\textbf{y}}}_{i+1})
\end{equation}
and the decoder is formulated as
\begin{equation}
 h_{\bW} (\mathbf{U}, \bZ, \mathbf{p}) = 
\sum_{i=1}^N \sum_{t_e=1}^{T_s} \sum_{t_s=0}^{t_e-1}  D_{i,t_e,t_s}
\end{equation}

Each feature function captures a distinct structural aspect of alignment quality, one is based on spectral dissimilarity at the local level, and the other measures contextual consistency at the global level. The feature functions used in the decoder are described next.

The first feature function, $\phi_1$, captures acoustic transition points in the $\bz$ domain.
We set the score for a boundary-candidate $\tilde{\textit{\textbf{y}}}_i $ in the frame $i$ to be the dissimilarity score between it and its consequent frame $i+1$.
\begin{equation}
    \phi_1 (\bZ,\tilde{\textit{\textbf{y}}_i}) =\frac{\partial s(\tilde{\textit{\textbf{y}}}_i,\tilde{\textit{\textbf{y}}}_i+1)}{\partial i} 
\end{equation}

\begin{figure}[t]
\centering
\includegraphics[width=\columnwidth ,height=0.9\columnwidth,keepaspectratio]{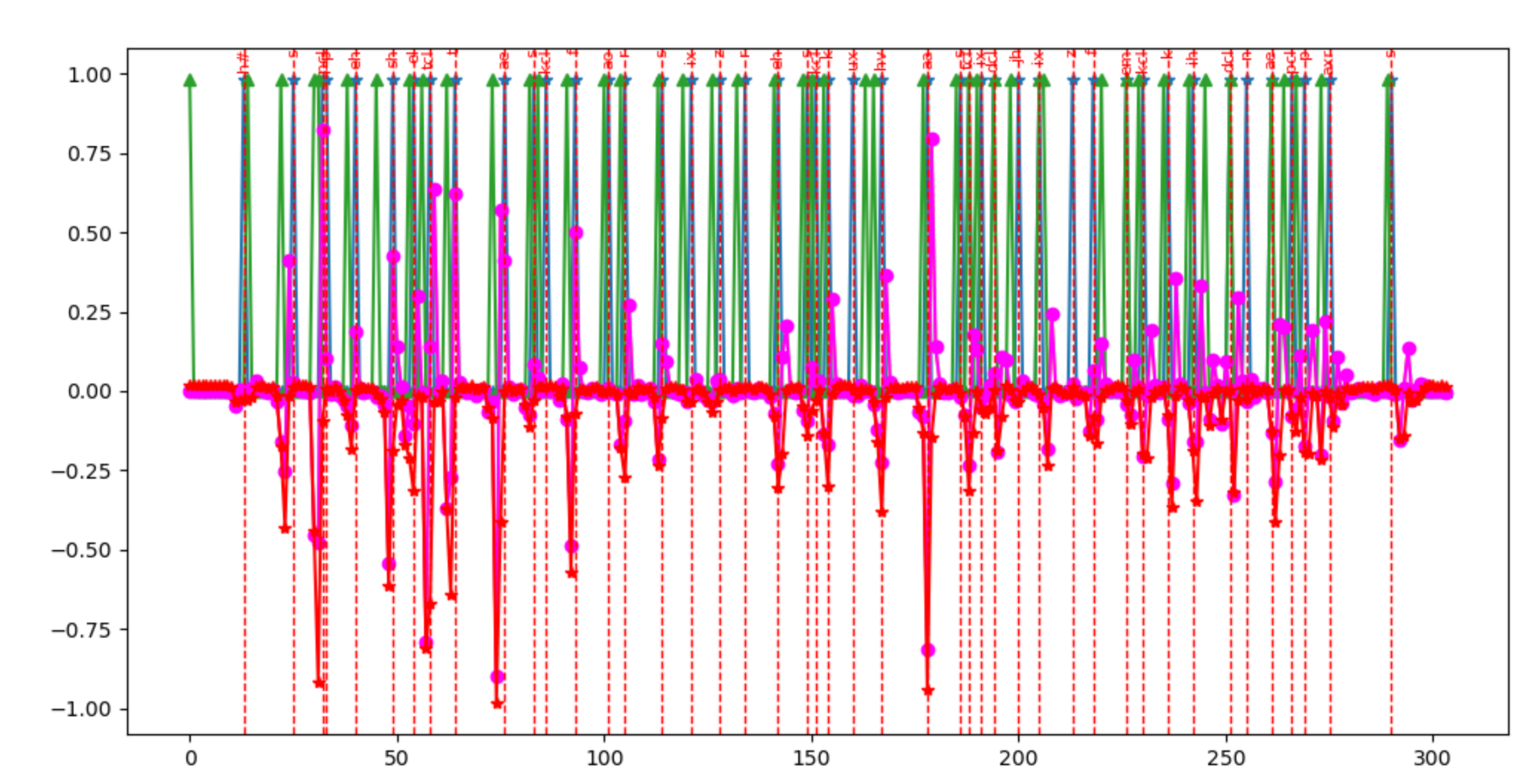}
\caption{Frame-wise cosine similarity (red) and its temporal derivative (pink) computed over the learned latent representation $\bz$, used by the decoder via feature $\phi_1$. Red dashed lines indicate ground-truth phoneme boundaries, and green shows the final alignment predicted by the full system. Peaks in the derivative correspond to likely phoneme boundaries.}
\label{fig:contrastive_graph}
\end{figure}

Where the similarity score function is the cosine similarity between consecutive samples in the $\bz$ domain
$s(\tilde{\textit{\textbf{y}}}_i,\tilde{\textit{\textbf{y}}}_i+1)$.
The frame-wise derivative of this score captures the temporal changes in spectral similarity. As shown in Fig. \ref{fig:contrastive_graph} High dissimilarity is trained to be represented at phoneme boundaries in large distances in the latent representation $\bZ$, resulting in local peaks, corresponding to likely phoneme transition points.


\begin{figure}[t]
\centering
\includegraphics[width=\columnwidth]{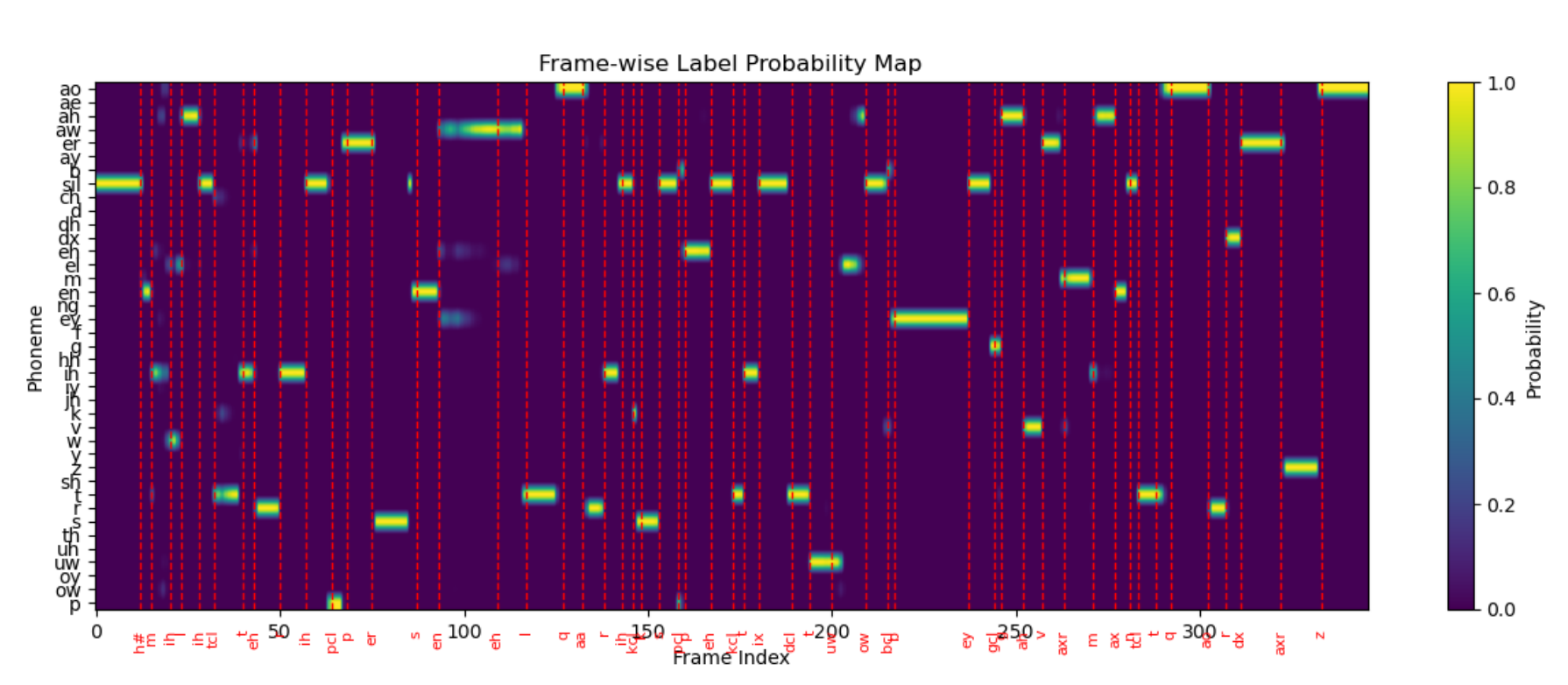}
\caption{Frame-wise phoneme probability map $\bU$ produced by the contextual encoder. The map encodes soft linguistic constraints used by the decoder via feature $\phi_2$. Red dashed lines indicate ground-truth phoneme boundaries, and the phoneme sequence $\bp$ is shown along the x-axis. The x-axis corresponds to latent frames, and the y-axis corresponds to phoneme classes $\mathcal{P}$}.
\label{fig:bilstm_probs}
\end{figure}

The second feature, $\phi_2$, measures the linguistic consistency of the alignment. Given the frame-wise phoneme probabilities, $g_{p_i}(\bz_j)\in \bU$ is the probability that at frame $\bz_j$ the phoneme label is $p_i$, as illustrated in Fig. \ref{fig:bilstm_probs}.
We define
\begin{equation}
    \phi_2(\bU,p_i,\tilde{\textit{\textbf{y}}}_{i},\tilde{\textit{\textbf{y}}}_{i+1}) = \frac{1}{\tilde{\textit{\textbf{y}}}_{i+1}-\tilde{\textit{\textbf{y}}}_{i}} \sum_{j=\tilde{\textit{\textbf{y}}}_{i}}^{\tilde{\textit{\textbf{y}}}_{i+1}} g_{p_i}(\bz_j)
\end{equation}

The feature represents the confidence of candidate boundaries for the phoneme $p_i$ by integrating probabilities over the candidates $\tilde{\textit{\textbf{y}}_i}, \tilde{\textit{\textbf{y}}}_{i+1}$. Normalizing by the segment length ensures that correct boundaries, which maximize the sum of the target phoneme's probability, receive higher scores, while preventing the model from overestimating the length of high-confidence segments.


Finding the alignment sequence that maximizes $h_{\bW}$ is done using a soft differentiable version of dynamic programming.
\begin{multline}
\hat{\mathbf{y}}^{*} = \arg{\max_{\hat{\mathbf{y}}}} \sum_{i=1}^N \sum_{\tilde{\textit{\textbf{y}}}_{i+1}=1}^{T_s} \sum_{\tilde{\textit{\textbf{y}}}_{i} = 0} ^{\tilde{\textit{\textbf{y}}}_{i+1}-1}  W_1  \phi_1(\bZ,\tilde{\textit{\textbf{y}}}_i) \\
+~W_2 \phi_2(\bU, p_i,\tilde{\textit{\textbf{y}}}_{i},\tilde{\textit{\textbf{y}}}_{i+1}).
\end{multline}

We further extend the DP decoding scheme to be fully differentiable with respect to the encoder outputs via a soft-DP formulation, inspired by the success of the soft-DTW \cite{cuturi2017soft} approach. The \emph{maximum} operation is replaced by a weighted sum over all possible alignment paths, with a differentiable weighting coefficient. Formally, denote $V$ as the DP table to be filled by the DP forward process, we define the soft recursion by replacing the hard \emph{maximum} over previous start times with the log-sum-exp operator scaled by a smoothing (temperature) parameter $\gamma$
{\small
\begin{multline}
V_{i,t_e,t_s} = D_{i,t_e,t_s} +
\gamma \log \sum_{t_{s_{prev}}=0}^{t_s-1}
\exp{\left({V_{i-1,t_s,t_{s_{prev}}}}/{\gamma}\right) }
\end{multline}
}
where the summation is taken over all valid candidate look-back start times $t_{s_{prev}} < t_s $ of the previous phoneme in the sequence $p_{i-1}$. The smoothing parameter $\gamma$ allows the operator to approximate the hard \emph{maximum} while maintaining a continuous gradient flow back to the encoders $f_\theta , g_\psi $.

\begin{figure}[t]
\centering
\includegraphics[width=\columnwidth ,height=0.9\columnwidth,keepaspectratio]{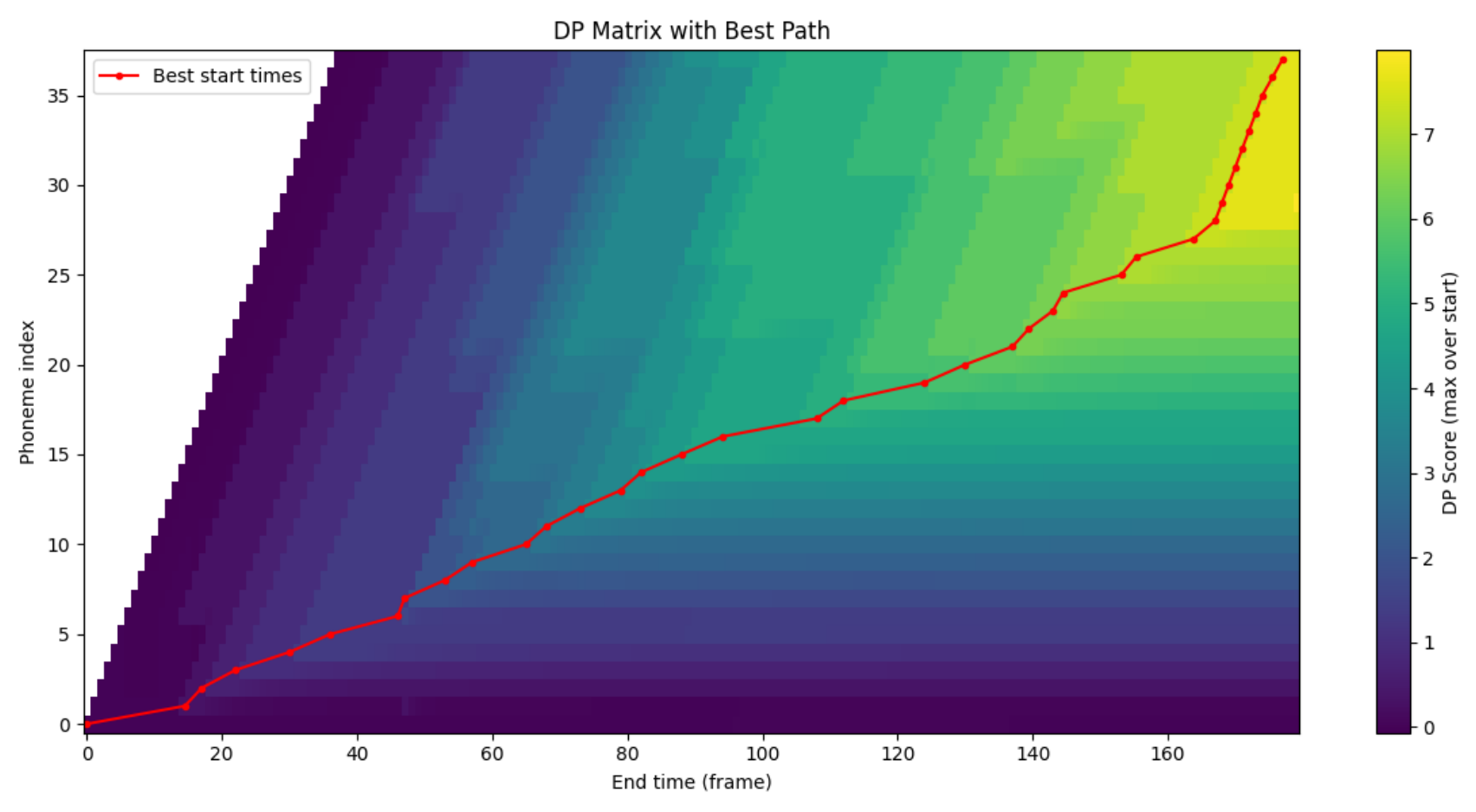}
\caption{Soft Dynamic Programming (Soft-DP) boundary selection. The matrix shows accumulated scores $V_{i,t_e, t_s}$ for candidate boundaries, and the highlighted path indicates the differentiable expected boundary estimation used during training and inference.}
\label{fig:soft_dp_matrix}
\end{figure}

Figure~\ref{fig:soft_dp_matrix} illustrates the soft-DP mechanism used to predict phoneme boundaries. Each matrix entry $V_{i,t_e,t_s}$ stores the accumulated score of candidate boundaries for phoneme $i$, computed using features $\phi_1$ and $\phi_2$. The optimal alignment is obtained as a differentiable expected path through the matrix, allowing gradient-based training of the encoder and context modules.

Moreover, standard backtracking via \emph{arg-max} operation is non-differentiable. Therefore, to maintain differentiability throughout the entire pipeline we denote the predicted boundary $\hat{y}_i$ as an expected value. Given the end-frame $t_e$ of the phoneme (which serves as the start-boundary for the subsequent phoneme), we compute the probability distribution over all candidate start-frames $t_s\in \{0,t_e-1\}$ using a softmax over the DP table $V$

\begin{equation}
\alpha_{i,t_e,t_s} = \frac{\exp({V_{i,t_e,t_s}}/{\gamma})}{\sum\limits_{t_s'=0}^{t_e-1} \exp({V_{i,t_e, t_s'}}/{\gamma})}
\end{equation}

The predicted boundary $\hat{y}_i$ is then extracted as the expectation
\begin{equation}
    \hat{y}_i = \sum_{t_s=0}^{T_s-1}\alpha_{i,t_e,t_s}\cdot t_s
\end{equation}

This allows the $l_2$ regression loss to be directly applied to the predicted timestamps $\hat{y}_i$.
while compared to $y_i$ the ground-truth boundary, aiming to predict the optimal $\hat{\by}_i^*$ during  training.

The full procedure is given in Algorithm \ref{alg:softdp}.

\begin{algorithm}
\caption{Soft-DP Phoneme Alignment}\label{alg:softdp}
\begin{algorithmic}
\STATE \textbf{Input:} 
Phoneme sequence $\bp$, features $\phi_1, \phi_2$, weights $w_1, w_2$, smoothing parameter $\gamma$.
\STATE \textbf{Output:} Expected phoneme boundaries  $\hat{\textbf{y}}$
\STATE \textbf{Initialization:}
$i=0, \forall t_e \in \{1, \dots, {T_s}\},t_s \in \{0, t_e-1\} $
\STATE\hspace{2cm} $V_{0, t_e, t_s} \gets D_{0, t_e, t_s}$
\STATE
\STATE \textbf{Forward Pass:}
\FOR{$i = 1, \dots, N$}
\FOR{each candidate end $t_e \in \{1, \dots, {T_s}\}$}\STATE 
Identify candidate starts $t_s \in \{0, t_e-1\}$\STATE 
Compute segment scores: 
\STATE
$t_{e_{prev}}\gets t_s$ \STATE
$D_{i,t_e,t_s} = w_1 \phi_1 + w_2 \phi_2$
\STATE $V_{i, t_e, t_s} \gets D_{i, t_e, t_s} ~ +$
\STATE\hspace{1cm} $\gamma \log \sum\limits_{t_{s_{prev}}=0}^{t_s-1} \exp({V_{i-1, t_{e_{prev}}, t_{s_{prev}}}}/{\gamma})$
\ENDFOR\ENDFOR\STATE\STATE \textbf{Backtracking (Expectation):}\STATE $t_e\gets {T_s}$\FOR{$i = N, \dots, 1$}\STATE $\alpha_{i,t_e,:} \gets \text{Softmax}({V_{i ,t_e, 0:t_e-1}}/{\gamma})$\STATE 
$\hat{y}_i \gets \sum\limits_{t_s=0}^{t_e-1} \alpha_{i,t_e,t_s} \cdot t_s$ \STATE $t_{e} \gets \text{round}(\hat{y}_i)$
\ENDFOR
\end{algorithmic}
\end{algorithm}

\section{Experiments}

In this section, we provide a detailed description of the experiments. We start by presenting the experimental setup, then we outline the experimental results and analysis.

\subsection{Datasets}
We trained and evaluated our model on two English speech corpora, TIMIT \cite{Garofolo93-TIM} and Buckeye \cite{Pitt05-BUC}, both of which provide high-quality speech recordings and corresponding phonetic and orthographic manually aligned timed transcription. TIMIT is a corpus of American English read speech which contains a total of 360 speakers and 6300 utterances that span 5.4 hours. The Buckeye Corpus of conversational speech spans 40 hours of hand-transcribed speech from 40 speakers of American English. For the TIMIT corpus, we used the standard train/test split, where we randomly sampled 10\% of the training set for validation. For Buckeye, we split the corpus at the speaker level into training, validation, and test sets with a ratio of 80/10/10, and we split long sequences into smaller ones by cutting during noise, silence, and untranscribed segments, following \cite{kreuk2020phoneme}.

For Dutch, we used a 10\% randomly sampled test set of the IFA Dutch Corpus \cite{VanSon01-IFA}, a database of approximately 5 hours of hand-segmented speech from 8 speakers, covering a range of speaking styles. For German, we evaluated on 10\% randomly sampled test set of the PHONDAT German Corpus \cite{tillmann1993theoretical}, which includes 201 speakers and 21587 utterances.
For Hebrew, we used a broadcast news dataset comprising 10 minutes of speech annotated at the phoneme level by professional linguists \cite{benshalom14}.

\subsection{Experimental Setup}
The function $f$ was implemented as a convolutional neural network, constructed of 5 blocks of 1-D strided convolution, followed by Batch-Normalization and Leaky ReLU \cite{maas2013rectifier} non-linear activation function. The network $f$ has kernel sizes of (10, 8, 4, 4, 4), strides of (5, 4, 2, 2, 2), and 256 channels per layer. Finally, the output was linearly projected by a fully connected layer. Overall the model was similar to the one proposed by \cite{Kreuk20-SCL}\cite{schneider2019wav2vec}.
This scheme serves as the latent representation $\bZ$ and also serves as an input to the function $g$ after normalization.

The contextual encoder $g$ was implemented as a 5-layer bidirectional Long-Short-Term-Memory (BiLSTM) network, where 5 was the minimal depth not causing mode-collapse on the validation set, with a hidden size of 512 and a fully connected projection layer to $|\mathcal{P}| =39$ classes according to \cite{46546}.

The decoder $h$ was implemented with $\gamma = 1e-20$, selected empirically from the range \{1e-1, 1e-3, 1e-4, 1e-20\}, as suggested in \cite{cuturi2017soft}.
Results were insensitive to moderate variations of this value. 


For the contrastive loss, we randomly sample 5 samples from $\cZ_\tau^{+}$ and $\cZ_\tau^{-}$ following \cite{Kreuk20-SCL}, and $\delta=\pm1$ frames. We experimented with $\delta \in \{1, 2, 4, 8 \}$ and observed that the temporal boundary's accuracy resolution increases as the boundary sampling window decreases.

We optimized the model with a batch size of 8 examples and a learning rate of 3e-4 for 200 epochs. We follow an early-stopping criterion computed over the validation set. All reported results are computed over the train set.
For the overall objective we set $\eta=2*10^{-9}$ and $\mu=10^{-4}$ as scaling constants used to balance the loss terms, based on their values on the TIMIT\cite{Garofolo93-TIM} dataset.

\subsection{Results}

We evaluated our suggested model, which will be reffered as \textit{FALCON: Forced Alignment through Contrastive Optimization Networks}, with the state-of-the-art baselines selected by \cite{Rousso24-FACOMP}: for the statistical approach, we compare its performance with the HMM-GMM approach tool, the Montreal Forced Aligner (MFA) \cite{McAuliffe17-MFA}, a widely addopted forced alignment system. For the neural forced aligners, we compare our performance with the Massively Multilingual Speech (MMS) model, which follows the wav2vec2.0 architecture \cite{wav2vec2} and was trained on over 1100 languages and approximately 500k hours of speech. \cite{pratap2024scaling}, and against WhisperX \cite{Bain23-WhisperX}, and Nvidia-Canary-1B\cite{nvidia-canary}.

We report FALCON both as a \textit{specialist} (trained and tested on the same corpus, i.e. TIMIT on TIMIT, Buckeye on Buckeye), and as a single \textit{joint} model trained on TIMIT and Buckeye combined. The \textit{specialist} captures in-domain performance, while the \textit{joint} model tests whether one shared model can match it without corpus-specific training. The primary multilingual results use the joint TIMIT+Buckeye checkpoint, transferred to the unseen languages.

We evaluated alignment accuracy for each model across multiple tolerance thresholds, consistent with the evaluation method in \cite{Rousso24-FACOMP}.

\medskip
\subsubsection{Phoneme-Level Alignment Performance on English}

In Table \ref{T1}, we compare our proposed model (FALCON) with the MFA baseline \cite{McAuliffe17-MFA} over phoneme-level FA. Note that it is the only aligner to provide the phoneme-level granularity among the proposed aligners.
Noteably, our proposed model surpasses the MFA across both TIMIT and Buckeye datasets in almost all evaluation thresholds.

\begin{table}[ht]
\renewcommand{\arraystretch}{1.3}\centering
\caption{Phone-level Alignment Accuracy [\%]: MFA vs. FALCON (Ours)}
\resizebox{\columnwidth}{!}{
\begin{tabular}{l | lc c c c}
\hline
& & \multicolumn{4}{l}{\textbf{alignment accuracy to ms threshold [\%]}}\\
\cline{3-6}
Dataset & Model & $t\leq10$ & $t\leq25$ & $t\leq50$ & $t\leq100$ \\
\hline
\multirow{3}{*}{TIMIT}
& MFA                      & \textbf{38.6} & 72.3 & 81.1 & 84.6 \\
& FALCON specialist   & 37.66 & \textbf{83.88} & \textbf{94.85} & \textbf{98.62} \\
& FALCON joint              & 34.70 & 82.62 & 94.91 & 98.60 \\
\hline 
\multirow{3}{*}{Buckeye}
& MFA                      & \textbf{35.3} & 60.6 & 68.9 & 72.7 \\
& FALCON specialist  & 29.69 & \textbf{69.93} & \textbf{90.07} & \textbf{97.40} \\
& FALCON joint              & 28.87 & 69.40 & 89.53 & 97.13 \\
\hline
\end{tabular}}
\label{T1}
\end{table}

These results demonstrate that our approach outperforms MFA across all practical tolerances (25-100 msec) on both datasets. Although MFA achieves higher accuracy at the strict 10 msec threshold, our method was not explicitly optimized for this granularity due to its 10 msec \emph{frame} resolution and a minimal boundary sampling window of $\pm1$ frames during training. 

\medskip

\subsubsection{Unseen Languages Generalization}
\label{unseen languages phoneme level}

We examined the robustness of our approach for multilingual generalization.
Although the FALCON alignment method was trained exclusively on English, we evaluated the model on the unseen languages Dutch, German, and Hebrew without further training. 
To address cross-lingual phoneme set variability, unseen-language phonemes were mapped to their closest IPA sequence of representations by articulatory feature distances according to PanPhon \cite{Mortensen-et-al:2016}, and then converted to the Lee-Hon39 \cite{46546} phoneme set employed in training.

\begin{table}[ht]
\renewcommand{\arraystretch}{1.3}\centering
\caption{Phoneme-Level: Unseen Multilingual Generalization Accuracy}
\resizebox{\columnwidth}{!}{
\begin{tabular}{l|l cccccc}
\hline
& & \multicolumn{6}{l}{\textbf{Alignment accuracy to ms threshold [\%]}}\\
\cline{3-8}
Test Set & Model & $\leq10$ & $\leq15$ & $\leq20$ & $\leq25$ & $\leq50$ & $\leq100$ \\
\hline
\multirow{3}{*}{Dutch - IFA}   & \textbf{FALCON joint}        & \textbf{26.85} & \textbf{36.16} & \textbf{44.56} & \textbf{51.17} & \textbf{69.94} & \textbf{84.11} \\
                               & FALCON specialist & 26.86 & 35.79 & 43.85 & 50.34 & 68.68 & 83.22 \\
                               & MFA                         & 11.01 & 14.70 & 19.05 & 21.80 & 33.90 & 51.02 \\
\hline
\multirow{3}{*}{German - PHONDAT} & \textbf{FALCON joint}        & \textbf{25.63} & \textbf{34.12} & \textbf{41.87} & \textbf{49.07} & \textbf{70.04} & \textbf{84.58} \\
                                  & FALCON specialist & 25.08 & 33.37 & 40.76 & 47.43 & 68.27 & 82.44 \\
                                  & MFA                         & 20.60 & 31.75 & 37.17 & 45.83 & 66.78 & 79.19 \\
\hline
\multirow{2}{*}{Hebrew} & \textbf{FALCON joint}        & \textbf{21.98} & \textbf{30.10} & \textbf{36.91} & \textbf{42.78} & \textbf{63.07} & \textbf{80.41} \\
                        & FALCON specialist & 21.03 & 27.78 & 34.30 & 39.79 & 59.38 & 77.76 \\
\hline
\end{tabular}}
\label{T2}
\end{table}

As shown in Table \ref{T2}, the proposed approach outperforms MFA across all alignment thresholds for Dutch and German, despite requiring no further training or language-specific models. In contrast, MFA relies on a separate acoustic model and dictionary for each language. For Hebrew, MFA cannot be evaluated due to the absence of a dictionary and acoustic model, highlighting that FALCON method is the only approach capable of performing phoneme-level alignment for this language. These results demonstrate that our model generalizes effectively to unseen languages without additional training.

\medskip

\subsubsection{Word-Level Alignment Generalization Evaluation}

e further evaluate the proposed FALCON model on the task of word-level forced alignment and compare its performance with both conventional statistical HMM-GMM-based aligners (MFA), which are also phoneme-based, and state-of-the-art neural forced alignment systems (MMS, WhisperX, and NVIDIA Canary-1B), which rely on CTC-based architectures and do not operate at the phoneme level.

The evaluation is first conducted on the English TIMIT and Buckeye corpora, with results reported in Table \ref{T3}, and subsequently extended to previously unseen multilingual datasets, as shown in Table \ref{T4}. Notably, no additional training or fine-tuning is performed for the word-level alignment task. The proposed model is trained exclusively on English phoneme-level annotations and is evaluated directly in a zero-shot manner across all datasets.

To ensure that the forced-alignment model is the only component being evaluated, we adopt MFA’s word-to-phoneme conversion pipeline without modification. Each orthographic word is converted into a phoneme sequence via pronunciation-dictionary lookup, while out-of-vocabulary words are handled using MFA’s Pynini \cite{gorman-2016-pynini} grapheme-to-phoneme (G2P) model. Optional inter-word silences are inserted according to MFA’s lexicon. The resulting phoneme sequence is then mapped to the Lee–Hon 39-phone inventory and aligned to the audio using the proposed FALCON model, with word boundaries derived from the predicted phoneme boundaries. Consequently, the proposed approach is identical to MFA up to the alignment stage. It should be noted that both FALCON and MFA employ the same G2P and lexical processing pipeline, whereas the neural alignment systems (MMS, WhisperX, and NVIDIA Canary-1B) operate directly at the word or token level without an explicit G2P stage. For this reason, the two groups of methods are separated by a horizontal dashed line in Tables \ref{T3} and \ref{T4}.

\begin{table}[ht]
\renewcommand{\arraystretch}{1.3}\centering
\caption{Word-Level Alignment Accuracy [\%]: Comparative Analysis}
\resizebox{\columnwidth}{!}{
\begin{tabular}{l | lcccc}
\hline
& & \multicolumn{4}{l}{\textbf{Alignment accuracy to ms threshold [\%]}}\\
\cline{3-6}
Dataset & Model & $t\leq10$ & $t\leq25$ & $t\leq50$ & $t\leq100$ \\
\hline
\multirow{6}{*}{TIMIT}
& FALCON spec (MFA-G2P)           & 49.22 & \textbf{81.79} & \textbf{93.04} & 98.37 \\
& FALCON joint (MFA-G2P)          & \textbf{49.50} & 80.60 & 92.86 & \textbf{98.46} \\
& MFA                            & 41.60 & 72.80 & 89.40 & 97.40 \\
\cdashline{2-6}
& MMS                            & 18.60 & 43.50 & 75.70 & 94.70 \\
& WhisperX                       & 22.40 & 52.70 & 82.40 & 94.20 \\
& Nvidia-Canary-1b               & 9.23 & 23.11 & 44.23 & 72.81 \\
\hline
\multirow{6}{*}{Buckeye}
& FALCON spec (MFA-G2P)           & 50.06 & 77.85 & \textbf{91.51} & \textbf{96.63} \\
& FALCON joint (MFA-G2P)          & \textbf{50.42} & \textbf{77.98} & 91.01 & 96.55 \\
& MFA                            & 39.80 & 69.90 & 84.90 & 91.80 \\
\cdashline{2-6}
& MMS                            & 25.00 & 52.70 & 75.00 & 87.90 \\
& WhisperX                       & 18.80 & 43.10 & 67.40 & 77.40 \\
& Nvidia-Canary-1b               & 8.06 & 18.83 & 36.31 & 63.29 \\
\hline
\end{tabular}}
\label{T3}
\end{table}

Table \ref{T3} reports word-level alignment accuracy on English TIMIT and Bcukeye datasets. We compare our model with both statistical and neural baselines, including  MFA, MMS, WhisperX and Nvidia-Canary-1B. A key distinction between these approaches lies in the prior knowledge required. MFA relies on both training and inference with pronunciation dictionaries and language-specific acoustic models, whereas neural baselines are trained or fine-tuned on large amounts of word-level data. Our model performs word-level alignment using only phoneme-level supervision, together with the same word-to-phoneme conversion at word-level inference only. Despite the absence of word-level training, the proposed FALCON aligner outperforms baselines in all reported tolerances. 

Multilingual word-level boundaries are derived at test time by mapping the word's phoneme predictions using articulatory feature distance, as described in sub-section \ref{unseen languages phoneme level}, without any further training.

\begin{table}[ht]\centering
\caption{Word-Level: Unseen Multilingual Generalization Accuracy}
\resizebox{\columnwidth}{!}{
\begin{tabular}{l | l c c c c}
\hline
\multicolumn{2}{c}{} & \multicolumn{4}{c}{\textbf{Alignment accuracy [\%]}} \\
\cmidrule(lr){3-6}
\textbf{Dataset} & \textbf{Model} & $t\leq10$ & $t\leq25$ & $t\leq50$ & $t\leq100$ \\
\hline
\multirow{3}{*}{German - PHONDAT}
  & FALCON (MFA-G2P)                 & \textbf{44.20} & \textbf{68.48} & \textbf{86.12} & \textbf{95.11} \\
  & MFA                             & 29.9 & 65.4 & 82.1 & 94.3 \\
    \cdashline{2-6}
  & MMS                             & 21.8 & 44.3 & 74.9 & 91.8 \\
\hline
\multirow{3}{*}{Dutch - IFA}
  & FALCON (MFA-G2P)                 &\textbf{ 26.38} & \textbf{45.15} & 61.16 & 76.49 \\
  & MFA                             & 4.7 & 7.3 & 11.6 & 19.0 \\
    \cdashline{2-6}
  & MMS                             & 16.0 & 37.9 & \textbf{62.9} & \textbf{76.6} \\
\hline

  \multirow{2}{*}{Hebrew}
    & FALCON
    & \textbf{31.91} &
  \textbf{56.72} & 75.18 & 87.89 \\
    \cdashline{2-6}
    & MMS
    & 14.3 & 41.3 & \textbf{76.5}
  & \textbf{94.7} \\
  

\hline
\end{tabular}}
\label{T4}
\end{table}

Table \ref{T4} presents the accuracy of the alignment of words at the word-level in unseen multilingual datasets. As in the phoneme-level multilingual evaluation \ref{unseen languages phoneme level}, the model is applied zero-shot, with no language-specific training. For German and Dutch, word boundaries are obtained through the same MFA word-to-phoneme conversion used for English, while out-of-vocabulary words are resolved with MFA's Pynini G2P model \cite{gorman-2016-pynini}. For Hebrew, MFA is unavailable due to the lack of a dedicated acoustic model and pronunciation dictionary. We therefore align at the word level without a G2P, mapping romanized characters directly to the Lee–Hon~39 inventory using articulatory-feature distance for generalization to unseen languages. Therefore, MMS is the only available baseline for Hebrew.
On German, our proposed model outperforms across all tolerances. On Dutch and German, it outperforms MFA by a large margin. On Dutch and Hebrew the proposed model exceeds MMS at
 the tighter tolerances ($t\leq25$~ms), and matches it in the other tolerances. These results indicate that the phoneme-trained model generalizes to word-level alignment across languages without additional training.
  

\leavevmode\\

\subsubsection{Architecture and Model Selection}


We next analyze the contribution of the main architectural components of our model, including the proposed MNCE loss and the soft dynamic programming decision module.

\begin{table}[t]
\renewcommand{\arraystretch}{1.3}
\centering
\caption{performance with MNCE vs. InfoNCE loss, same architecture on TIMIT}\label{tab:Ablations loss}
\resizebox{\columnwidth}{!}{
\begin{tabular}{lcccccc}
\hline
& \multicolumn{6}{c}{\textbf{Alignment accuracy to ms threshold [\%]}}\\
\cline{2-7}
loss  & $t \leq 10$ & $t \leq 15$ & $t \leq 20$ & $t \leq 25$ & $t \leq 50$ & $t \leq 100$ \\
\hline
MNCE$_{\text{ours}}$  & \textbf{37.66} & \textbf{57.99} & \textbf{74.7} & \textbf{83.88} & \textbf{94.85} & \textbf{98.62} \\
InfoNCE$_{\text{CPC}}$  & 21.5 & 27.62 & 32.46 & 36.93 & 53.65 & 72.51 \\
\hline
\end{tabular}
}
\label{T5}
\end{table}

Table \ref{T5} compares the proposed MNCE objective with the commonly used InfoNCE loss from CPC. Replacing InfoNCE with the suggested objective leads to substantial improvements across all alignment thresholds. These results indicate that the proposed loss formulation provides a suitable training representation for phoneme boundary detection.

\begin{table}[ht]
\renewcommand{\arraystretch}{1.3}
\centering
\caption{Ablation Study for Decision Module on TIMIT}
\label{tab:mfa_phone}
\resizebox{\columnwidth}{!}{
\begin{tabular}{llcccc}
\hline
& & \multicolumn{4}{c}{\textbf{Alignment accuracy to ms threshold [\%]}}\\
\cline{3-6}
Decision method & Bi-LSTM & $t \leq 10$ & $t \leq 25$ & $t \leq 50$ & $t \leq 100$ \\
\hline
Hard DP & No  & 20.9 & 40.2  & 59.05 & 77.79 \\
Hard DP & Yes & 27.9 & 49.06 & 66.11 & 82.38 \\
Naive Peak Detection & No & 24.7 & 62.0 & 98.0 & 99.0 \\
\textbf{Soft DP (ours)} & \textbf{Yes} & \textbf{37.66} & \textbf{83.88} &\textbf{94.85} & \textbf{98.62} \\
\hline
\end{tabular}
}
\label{T6}
\end{table}

Table \ref{T6} demonstrates the evaluation of different boundary decision approaches. Using Hard Dynamic programming already improves over naive peak detection, and incorporating the contextual feature further increases performance. However, the proposed soft DP decision module achieves the highest accuracy across all thresholds by a large margin. This demonstrates that combining a contextual representation with a contrastive one, and specifically using a fully differentiable decision module that enables joint training of both the decoder and the encoder, is very effective for accurate boundary estimation.

\section{Discussion}

The experimental results demonstrate that the proposed approach effectively bridges the gap between traditional statistical forced alignment systems and modern neural architectures. While end-to-end ASR-based models can produce accurate transcriptions and estimate word-level timestamps, they are not optimized for precise boundary estimation and therefore often exhibit limited frame-level temporal precision. In contrast, the proposed model is explicitly designed for phoneme boundary detection and, to the best of our knowledge, represents the first neural forced alignment system operating at the phoneme-level. Existing neural approaches, such as MMS, WhisperX, and NVIDIA Canary, primarily provide word-level timestamps, whereas the proposed method produces frame-level phoneme alignments.

The results indicate that combining contrastive representation learning with contextual phoneme modeling significantly improves boundary detection accuracy. The representation encoder learns boundary-sensitive acoustic features via the proposed objective, which emphasizes the distinction between intra-phoneme and phoneme-transition regions. The contextual encoder then aggregates information across neighboring frames to produce temporally consistent phoneme posterior probabilities. Together, these representations allow the decoder to evaluate candidate boundaries using both local spectral transitions and global linguistic consistency.

A key contribution of this work is the integration of differentiable soft dynamic programming (Soft-DP) within the alignment decision module. Accurate phoneme boundary detection remains challenging due to gradual acoustic transitions between adjacent phonemes and variability in phoneme durations. Unlike traditional peak detection or standard DP decoding, which in some cases is applied only during inference, the Soft-DP formulation enables gradient propagation through the alignment process. This allows the encoders to be optimized directly with respect to alignment quality. The ablation results confirm that this joint optimization improves alignment accuracy compared to commonly used decision methods that are not part of the training process.

While the ability to train all components end-to-end is advantageous, incorporating dynamic programming computations inside the training loop increases training time compared to standard neural architectures. This overhead results from the additional DP computations required during optimization. However, the inference complexity remains comparable to classic HMM-GMM-based aligners such as MFA, since both approaches rely on dynamic programming during inference.

Another notable property of the proposed method is its strong cross-lingual generalization capability. Despite being trained exclusively on English phoneme-level data, the model performs competitively on Dutch and German and can produce phoneme-level alignments for Hebrew without additional training. In contrast, traditional systems such as MFA require language-specific acoustic models and pronunciation dictionaries for both training and inference. These results suggest that the model captures universal acoustic cues associated with phoneme transitions rather than relying solely on language-specific phonetic representations, enabling effective transfer across languages.

Although the proposed method achieves strong performance across most alignment tolerance thresholds, it is sometimes constrained by the temporal resolution of the latent representation at extremely strict thresholds (e.g., 10 ms). The current architecture produces frame-level features at approximately 10 ms resolution, which limits the achievable boundary precision. Future work may therefore explore finer-grained acoustic representations or multi-scale encoders to improve temporal resolution.

Several additional research directions emerge from this work. First, training efficiency could be improved by developing more computationally efficient differentiable DP formulations or by constraining the DP search space using prior knowledge of phoneme duration distributions. Second, although the model is trained only at the phoneme level, it achieves competitive word-level alignment accuracy. Across many tolerance thresholds, particularly the strictest ones, the model outperforms existing neural aligners, including those that operate exclusively at the word level. However, in some tolerances, it matches the performance of the existing methods. Future work could therefore extend the proposed method to optimize phoneme and phoneme-to-grapheme mapping as an end-to-end objective jointly. Third, integrating large-scale self-supervised speech representations may further improve robustness and cross-lingual generalization.

In summary, the proposed system introduces a fully differentiable neural architecture for phoneme-level forced alignment. The results demonstrate that the approach achieves high alignment accuracy while maintaining strong cross-lingual generalization, without requiring large-scale training data. This property makes the method particularly attractive for low-resource languages where alignment tools and annotated data are limited.

\section{Acknowledgments}
This work was supported by NSF DRL Grant No. 2219843 and BSF Grant No. 2022618. We also thank Rob van Son for his guidance and support with the IFA Corpus.

\bibliographystyle{IEEEtran}  
\bibliography{references}           

\end{document}